\documentclass[aps,prl,showpacs,twocolumn,a4paper,10pt]{revtex4}
\usepackage{tipa}
\usepackage{amssymb}
\usepackage{amsmath}
\usepackage{graphicx}
\usepackage{dcolumn}
\usepackage{bm}

\begin{document}

\title{Quantum Blockade and Loop Currents in Graphene with Topological
Defects}
\author{Yan-Yang Zhang$^{1,2}$, Jiang-Ping Hu$^3$, B.A. Bernevig$^4$, X. R. Wang$^2$, X. C. Xie$^{5,1}$, and W. M. Liu$^1$}
\affiliation{$^1$Beijing National Laboratory for Condensed Matter
Physics, Institute of Physics, Chinese Academy of Sciences, Beijing
100080, China.} \affiliation{$^2$ Physics Department, The Hong Kong
University of Science and Technology, Clear Water Bay, Hong Kong
SAR, China} \affiliation{$^3$Department of Physics, Purdue
University, West Lafayette, Indiana 47907, USA}
\affiliation{$^4$Princeton Center for Theoretical Physics and
Department of Physics, Jadwin Hall, Princeton University, Princeton,
NJ 08544} \affiliation{$^5$Department of Physics, Oklahoma State
University, Stillwater, Oklahoma 74078, USA}
\date{\today}

\begin{abstract}We investigate the effect of topological defects on the transport properties
of a narrow ballistic ribbon of graphene with zigzag edges. Our
results show that the longitudinal conductance vanishes at several
discrete Fermi energies where the system develops loop orbital
electric currents with certain chirality. The chirality depends on
the direction of the applied bias voltage and the sign of the local
curvature created by the topological defects. This novel quantum
localization phenomenon provides a new way to generate a magnetic
moment by an external electric field, which can prove useful in
nanotronics.
\end{abstract}


\maketitle

Graphene is a single layer of graphite with a honeycomb lattice
consisting of two triangular sub-lattices. This peculiar structure
of graphene gives rise to two linear ``Dirac-like'' energy
dispersion spectra around two degenerate and inequivalent points $K$
and $K'$ at the corner of the Brillouin zone \cite{Gon93,Kane05}.
The valley index that distinguishes the two Dirac points is a good
quantum number, even in the presence of weak disorder, since
inter-valley scattering requires the exchange of large momentum.
Valley-dependent physics has been actively explored recently and can
potentially play an important role in future graphene based devices
\cite{Ryc07,Akh07,Niu07}.

\begin{figure} [t]
\begin{center}
\includegraphics[width=0.45\textwidth]{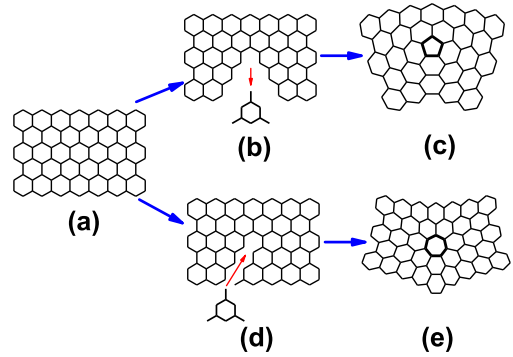}
\end{center}
\caption{(color online) The process of constructing a pentagon
((a)$\rightarrow$(b) $\rightarrow$(c)) and a heptagon
((a)$\rightarrow$(d)$\rightarrow$(e)) by ``cut'' and ``paste'' from
a perfect graphene sheet. Positive (negative) curvature is induced
by a pentagon (heptagon) in the realistic three dimensional space.
Notice that after the ``cut'' and ``paste'', a single pentagon
(heptagon) will give rise to a global change of the lattice
structure, which is explained in the text. }\label{Geometry}
\end{figure}

To manipulate this extra degree of freedom, it is necessary to
couple the two valleys. In graphene, there is a natural way to
produce a valley coupling by creating  topological defects such as
pentagons and heptagons \cite{An01}. Such defects cannot be
constructed from a perfect graphene sheet simply by replacing a
hexagon by a pentagon or a heptagon. Instead, a ``cut-and-paste''
process \cite{Cor06,Sit07} should be employed to keep the local
coordination number of each carbon atom, as illustrated in Fig.
\ref{Geometry}. A pentagon (heptagon) will induce a positive
(negative) curvature around it. As can be observed from Fig.
\ref{Geometry}, after going around any closed carbon loop encircling
the defect, which always consists of odd number of atoms, the roles
of two sublattices are interchanged. Therefore, the defect breaks
the bipartite nature of the lattice in the real space as well as the
symmetry between the $K$ and $K'$ points in reciprocal space,
leading to a M\"{o}bius strip-like structure coupling the two
valleys. Theoretically, the effect of topological defects on the low
energy electric physics of graphene is equivalent to generating
non-Abelian gauge potentials with the internal gauge group involving
the transformation of valley index. The wave function acquires a
topological phase when circling around the defect, which can be
described by means of a non-Abelian gauge field. Experimentally,
pentagon and heptagon topological defects have been found in
graphite-related materials \cite{An01,Jas03}. In graphene, recently
observed \cite{Nov04,Moro06} mesoscopic corrugations (ripples) are
partially attributed to topological defects.

The equilibrium electronic properties of a two dimensional graphene
in the presence of single or many
 topological defects have already received wide attention \cite{Cor06,Sit07,Lam00,Gon93,Char01,Ko00,Ta94,Ta97}.
However, there is still much to understand about the transport
properties of these systems. In this paper, we investigate the
electronic transport properties of a zigzag edge graphene nanoribbon
with several topological defects: a pentagon, a heptagon, and a
pentagon-heptagon pair at the center. We numerically calculate the
total conductance $G$ and the spatial distribution of local currents
$i(\mathbf{r})$.  We reveal a novel quantum localization phenomenon
whereby the conductance vanishes at discrete Fermi energies in the
first quantized plateau. This effect is accompanied by the
development of circular loop currents with prescribed chirality,
owing their existence to the non-Abelian gauge potentials connecting
the valleys. The chirality depends on both the direction of the
applied bias voltage and the sign of curvature created by
topological defects, and can hence be easily controlled and
manipulated. The result opens a new possibility to generate magnetic
moments by an external electric field in graphene.

The $\pi$ electrons in graphene are modeled by the tight binding
spinless Hamiltonian
\begin{equation}
H_G=\sum_{i}\epsilon_{0}c_{i}^{\dagger }c_{i}+t\sum_{\langle
i,j\rangle }(c_{i}^{\dagger }c_{j}+\text{H.c}),  \label{1}
\end{equation}%
where $c_{i}^{\dag }$ ($c_{i}$) creates (annihilates) an electron on
site $i$, $\varepsilon _{0}$ (set to 0) is the on-site energy and
$t$ ($\sim $2.7eV) is the hopping integral between the nearest
neighbor carbon atoms separated by $a$ ($\sim $1.42{\AA }). $t$ and
$a$ are used as the energy unit and the length unit, respectively.
The hopping $t$ is taken to be constant for all the C-C bonds, even
those on the defect. Small changes of the hopping integral $t$ due
to orbital mixing from the curvature \cite{Klein01} are neglected
since they do not generate large momentum scattering, which means
that pure topological effects are considered.

The zero temperature two terminal conductance $G$ of the sample and
local density of states (LDOS) $\rho_m$ of site $m$ at Fermi energy
$E_F$ are given by \cite{Datta}
\begin{eqnarray}
G(E_{F}) &=& \frac{2e^{2}}{h}\text{Tr}(\Gamma _{S}(E_{F})
G^{r}(E_{F})\Gamma _{D}(E_{F})G^{a}(E_{F})),\label{2} \\
\rho_m(E_{F}) &=& -\frac{1}{\pi }\text{Im}[G_{m,m}^{r}(E_{F})],
\label{3}
\end{eqnarray}
where $G^{r(a)}(E_{F})=(E_{F}-H_G-\Sigma _{S}^{r(a)}(E_{F})-\Sigma
_{D}^{r(a)}(E_{F}))^{-1}$ is the retarded (advanced) Green's
function, $G^{a}(E_{F})=(G^{r}(E_{F}))^{\dagger }$, $\Gamma
_{S(D)}(E_{F})=i[\Sigma _{S(D)}^{r}(E_{F})-\Sigma
_{S(D)}^{a}(E_{F})]$, $H_G$ is the Hamiltonian of the graphene
sample and $\Sigma _{S}^{r(a)}\equiv H_{GS}G^{r(a)}_S
H_{SG}=H_{GS}(E_{F}\pm i\eta-H_S)^{-1} H_{SG}$ is the retarded
(advanced) self-energy due to the semi-infinite source, and the
self-energy due to the drain $\Sigma _{D}^{r(a)}$ is similar. The
self-energies can be numerically calculated in a recursive way
\cite{Lee,JZhang}. We attach clean graphene as source and drain
leads to avoid redundant scattering from mismatched interfaces
between the sample and the leads.

The current per unit energy at the Fermi level $E_F$ between
neighboring sites $m$ and $n$  is \cite{Datta,Naka01,Anan07,Guo04}
\begin{eqnarray}\setlength\arraycolsep{2pt}
i_{m\rightarrow
n}(E_F)&=&\frac{4e}{\hbar}\langle\text{Im}[\psi_m^{*}t\psi_n]\rangle=\frac{4e}{\hbar}\langle H_{mn}|\psi_m| |\psi_n|\sin(\theta_n-\theta_m)\rangle\label{8}\\
&=&\frac{4e}{h}\text{Im}[H_{mn}G^{n}_{n,m}(E_F)], \label{9}
\end{eqnarray}
where $\psi_m$ is the amplitude of the wave function at site $m$,
$G^n\equiv G^R(\Gamma_S f(\mu_S)+\Gamma_D f(\mu_D)) G^A$ is the
electron correlation function and $f$ is the Fermi distribution
function, $\theta_n$ is the phase of $\psi_n$ and $H_{mn}\equiv t$.
The quantum current $i$ contains information about the phase shift
of wave functions between the neighboring sites.

The local current $I_{m\rightarrow n}$ is the energy integral of
$i_{m\rightarrow n}$:
\begin{eqnarray}
I_{m\rightarrow n} &=&\int^{\mu_S}_{-\infty} i_{m\rightarrow n}(E)dE\\
&=&\int^{\mu_S}_{\mu_D} \frac{4et}{h}\text{Im}[G^R(E)\Gamma_S(E) G^A(E)]_{mn}dE  \quad \textrm{\Big(zero temperature and zero magnetic field\Big)} \label{5}\\
&=&i_{m\rightarrow n}(\mu_S)eV_b \quad \textrm{\Big(linear response
limit \Big)} \label{6}
\end{eqnarray}
where the bias $V_b$ is related to the source (drain) chemical
potential $\mu_S$ ($\mu_D$) by $\mu_S-\mu_D=eV_b$. Equation
(\ref{5}) can be derived from equation (\ref{9}) in a
straightforward way along with the fact that Green's functions are
symmetric matrices in the absence of magnetic field and thermal
fluctuations, thus the contributions below $\mu_D$ cancel, making
the local current a Fermi surface property \cite{Datta}.

\begin{figure} [t]
\begin{center}
\includegraphics[width=0.5\textwidth]{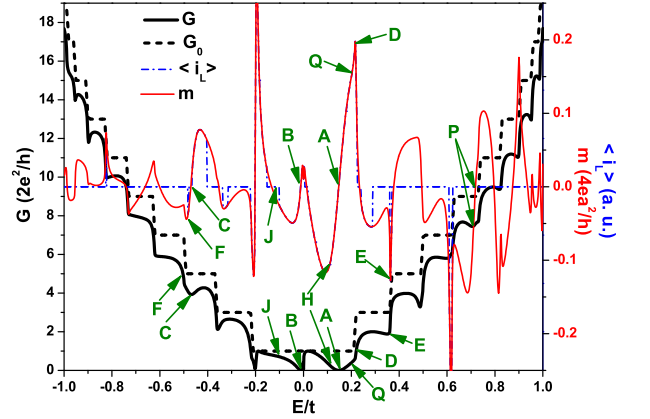}
\end{center}
\caption{(color online) Conductance of the topologically disordered
sample with a pentagon $G$ (black solid), averaged loop current
$\langle
 i_{L}\rangle$ (blue dashed dot) and induced magnetic moment $m$ (red solid) of the
pentagon as functions of Fermi energy $E$. The conductance for a
perfect graphene sample $G_0$ (black dashed) with the same
dimensions is also plotted for reference.} \label{PenEG}
\end{figure}

We constructed a pentagonal defect in the center of a zigzag edge
graphene sheet with $45\times20$ sites. in the presence of the
pentagon, as shown in Fig. \ref{PenEG}, the conductance $G$ is
suppressed by defect scattering in most regions of energy compared
to the quantized conductance \cite{Peres06} of the ballistic case.
The $G(E)$ curves exhibit dips and kinks at some special energy
points. The details of these points are the main focus of this work.

Let us first concentrate on the dips A and B in Fig. \ref{PenEG},
where the conductance is reduced by the amount of $g_0= 2e^2/h$,
suggesting a single conducting channel at this energy has been
completely blocked. The local shape of the dips is approximately
Lorentzian. An anti-resonance with such features has been observed
in one-dimensional (1D) systems with topological defects
\cite{Louie00,Guin87,Wang02}. The location of these anti-resonances
is not universal, but is determined by the topological configuration
of the lattice and its defects. Unlike the 1D case, the locations of
these anti-resonances do not have a simple relation with the energy
levels of the defect or the sample. The multi-dimensional nature of
the transmission matrix and the inter-channel scattering make an
analytical treatment more complicated. However, the physical result
is the same as that in 1D. The phenomenon at hand can be called
topological quantum localization.

\begin{figure} [t]
\begin{center}
\fbox{\includegraphics[scale=0.3]{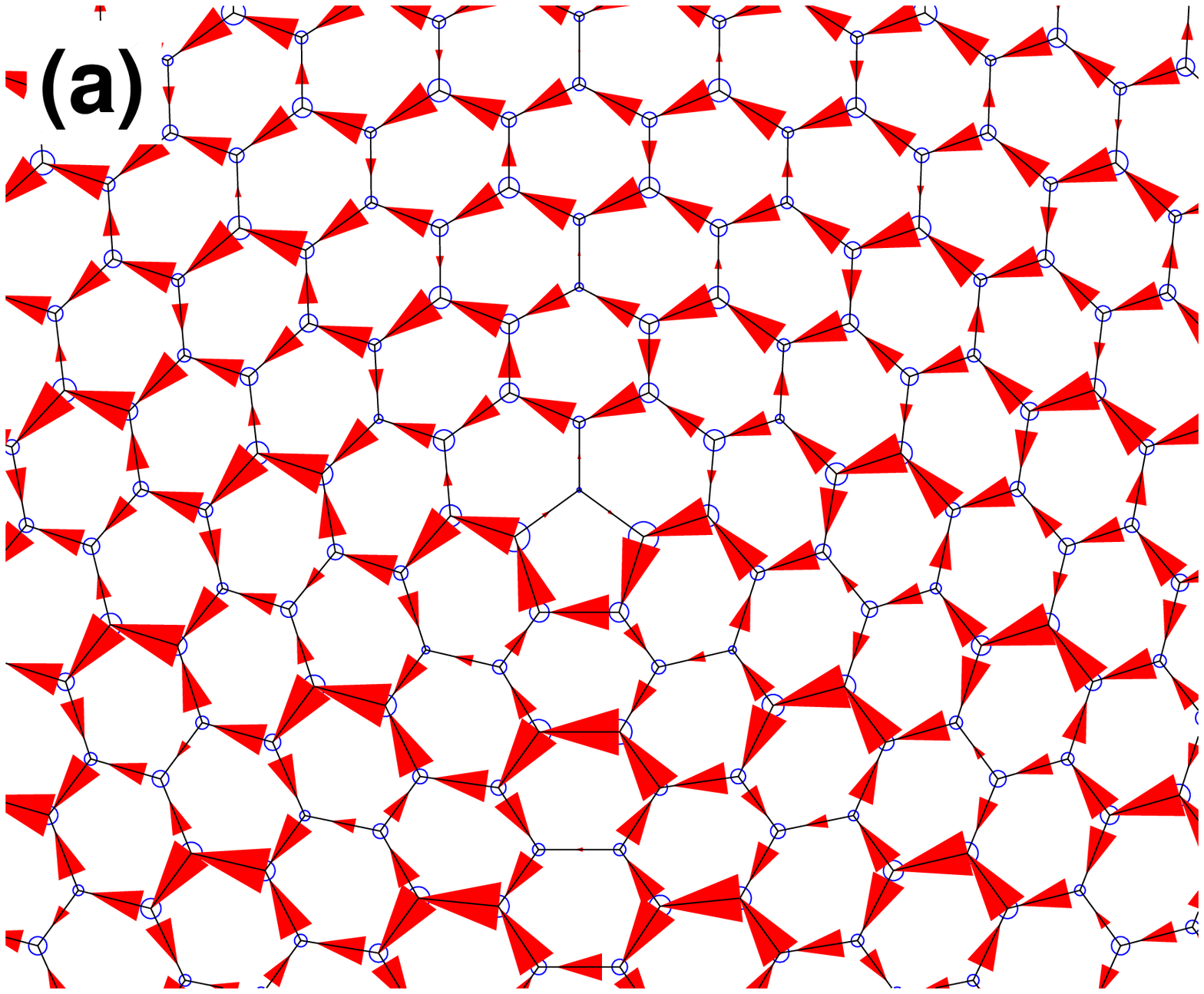}}
\fbox{\includegraphics[scale=0.3]{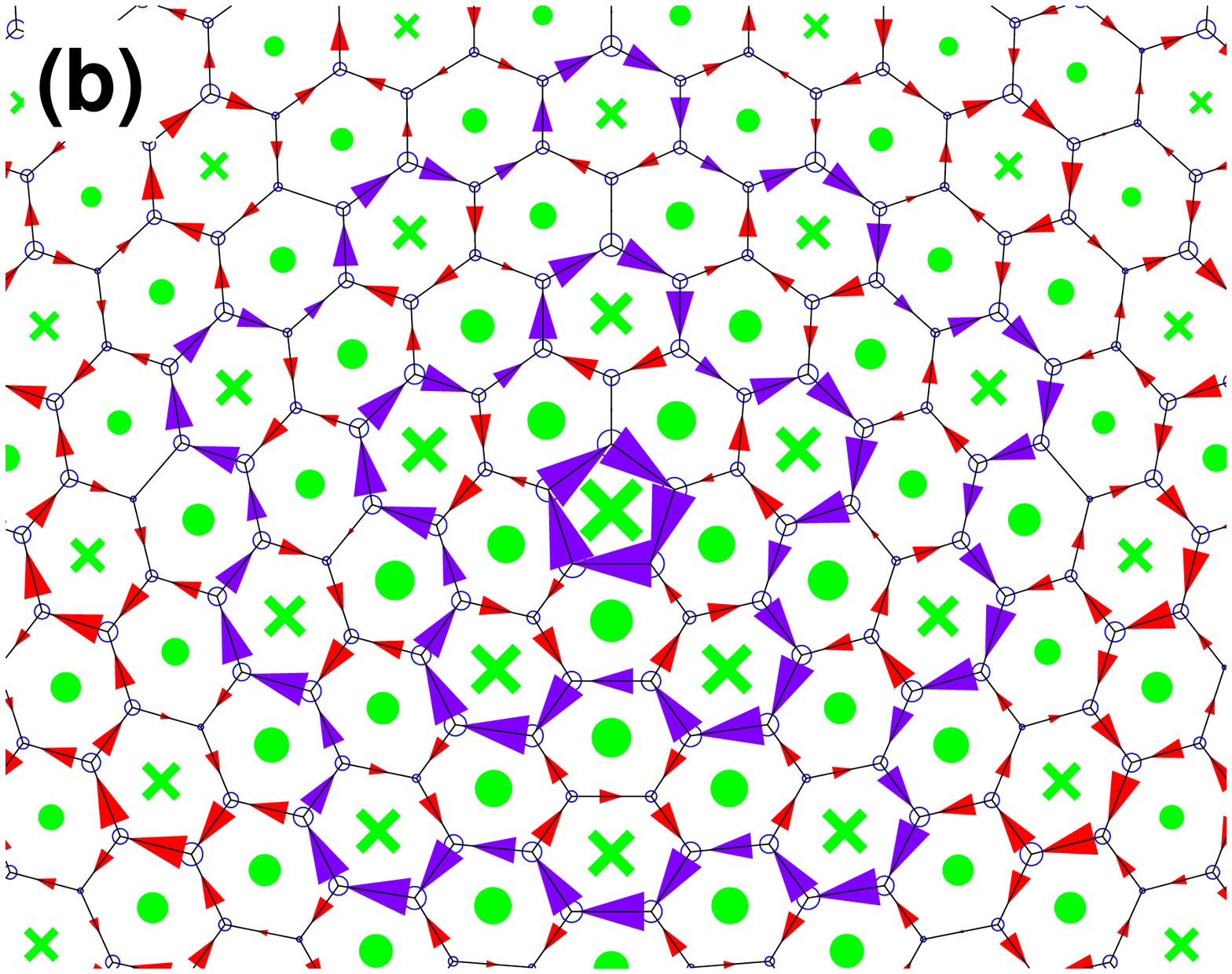}}
\fbox{\includegraphics[scale=0.3]{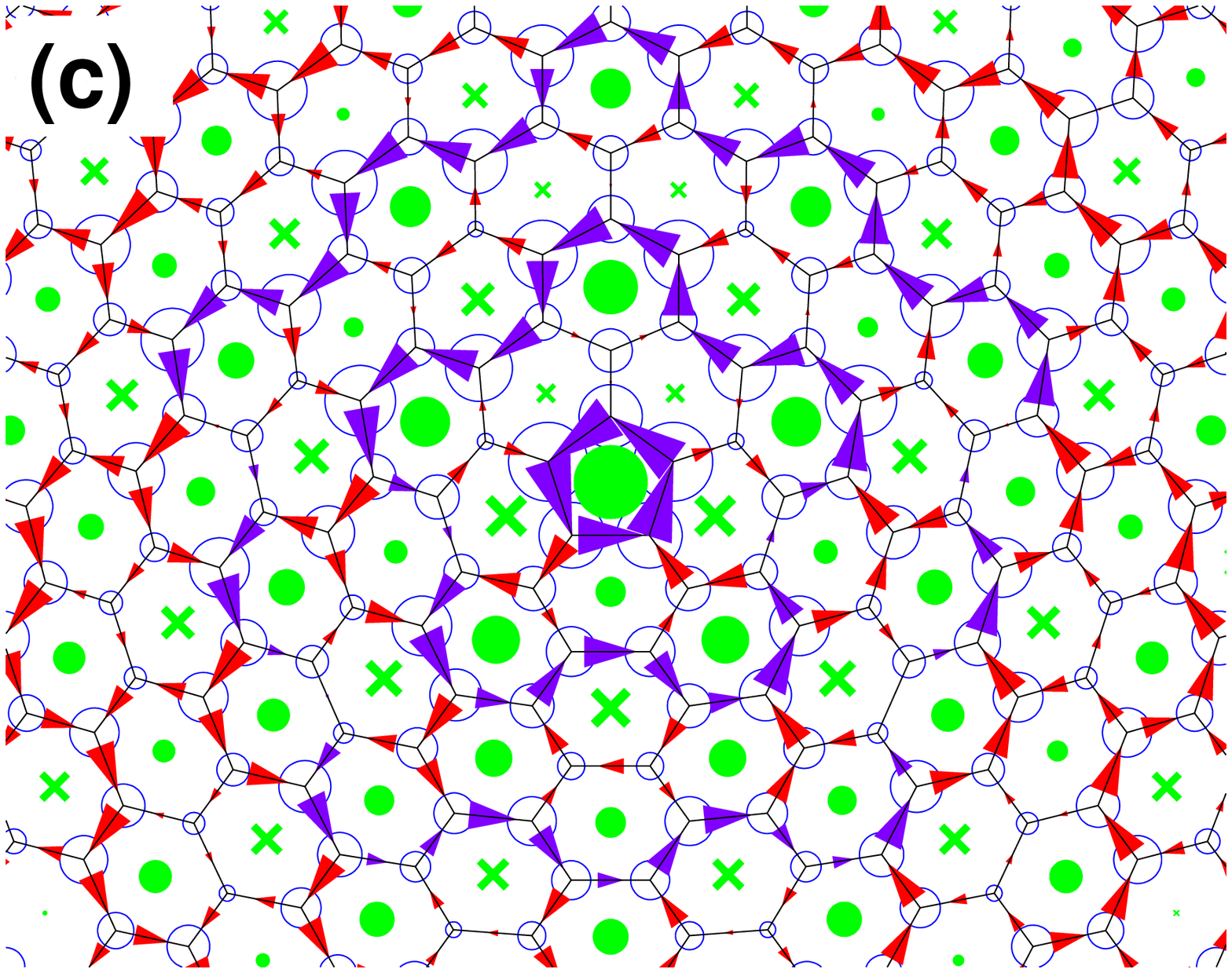}}
\end{center}
\caption{(color online) The spatial distribution of LDOS of the
sites (blue circles), local current density of the bonds (arrows)
and induced magnetic moment of carbon rings (green disks and
crosses) near the pentagon. The size of each symbol is proportional
to the local value of the corresponding quantity. (a) $E=-0.1t$,
corresponding to point J in Fig. \ref{PenEG}, far away from the
anti-resonance. (b) $E=0.1t$, corresponding to point H in Fig.
\ref{PenEG}, near the anti-resonance. (c) $E=0.2t$, corresponding to
point Q in Fig. \ref{PenEG}, on the other side of the
anti-resonance. Some circular loop currents around the pentagon are
stressed in purple arrows. Note that circular loop currents can also
be observed on some hexagons near the defect.} \label{Pentagon}
\end{figure}

To gain better insight in the microscopic origins of the
localization, we plot, in Fig. \ref{Pentagon} (a) and (b), the
spatial distribution of LDOS $\rho$, and local currents $i$ far
(point J in Fig. \ref{PenEG}) away from and near (point H in Fig.
\ref{PenEG}) the antiresonance respectively. The LDOS does not vary
much at the two points, but the currents do. In Fig. \ref{Pentagon}
(a), where $G$ is not sharply suppressed, the spatial distribution
of $i(\mathbf{r})$ is rather uniform, and almost all of the bond
currents flow from the source (right) to the drain (left). On the
contrary, in Fig. \ref{Pentagon} (b), loop currents flow circularly
around the defect in the same direction, giving rise to a
vortex-like pattern of local currents. Interestingly, these loop
currents reverse their chirality when the Fermi energy is swept from
one side to the other side of the antiresonance, as can be seen in
Fig. \ref{Pentagon} (c). The magnitude of the current on the
pentagon bonds may be larger than the source-drain current.
Moreover, loop currents on some of the hexagons in the proximity of
the defect can also be observed. This is a quantum effect, which is
forbidden in a classical resistance network \cite{Altshuler}, due to
the lack of an external battery or magnetic field on the loops. This
pattern exhibits almost perfect five-fold rotation symmetry of the
pentagon. Since the graphene sample used here lacks the rotational
symmetry of the pentagon, the pattern in Fig. \ref{Pentagon} (b)
does not originate from an artificial symmetric arrangement of the
device, but is an intrinsic property of the topological defect, and
can be expected to exist in the realistic system. In the case of an
infinite graphene sheet, the first conductance plateau containing
this bounding state encircling the defect degenerates into the Dirac
point. The existence of this bounding state at Dirac point is
consistent with previous predictions based on the the continuum
description of 2D graphene \cite{Cor06,Lam00,Sit07}.

The vortex-like loop currents suggest that the topological defect
introduces an effective magnetic field \cite{Wakabayashi}, which is
consistent with theoretical predictions \cite{Lam00,Cor06,Sit07}.
According to (\ref{9}), the loop currents reflect the peculiar phase
correlation induced by the topological defect. Given a closed carbon
loop, we define the bond current on the loop as positive if it
points clockwise, and negative otherwise. Thus the averaged loop
current (ALC) density $\langle i_{L}\rangle_{\text{loop}}$ of a
closed loop can be defined as
\begin{equation}
\langle
i_{L}\rangle_{\text{loop}}=\frac{1}{N}\theta(i_{\text{max}}\cdot
i_{\text{min}})\cdot\sum_{\langle k,l\rangle\in
\text{loop}}i_{k\rightarrow l},
 \label{67}
\end{equation}
where $N$ is the number of bonds (also the number of sites) on the
loop, $\theta(x)$ is the Heaviside step function, and
$i_{\text{max}}$ ($i_{\text{min}}$) is the maximum (minimum) current
on the loop. Obviously a current on the loop can be called
``circular'' only if $i_{\text{max}}\cdot i_{\text{min}}>0$, so that
$\langle i_{L}\rangle_{\text{loop}}\neq 0$. The induced magnetic
moment density on a closed loop is expressed as \cite{Naka01}
\begin{equation}\setlength\arraycolsep{2pt}
\mathbf{m}_{\text{loop}}(E)=\sum_{\langle k,l\rangle\in
\text{loop}}i_{k\rightarrow l}(E)(\mathbf{r}_k\times
\mathbf{r}_l)/2, \label{10}
\end{equation}
where $\mathbf{r}_k$ is the coordinates of the site $k$. A relation
similar to that of $I$ and $i$ in (\ref{5} and \ref{6}) holds
between the magnetic moment $\mathbf{M}$ and its energy density
$\mathbf{m}$. Due to equations (\ref{5}) and (\ref{6}), we simply
discuss the densities $i$ and $\mathbf{m}$ in the following, but
refer to them simply as current and magnetic moment. The ALC and
magnetic moment as functions of energy are plotted in Fig.
\ref{PenEG}. They change signs when energy crosses an anti-resonance
(A and B in Fig. \ref{PenEG}). This is an important correlation
between the magnetic moment direction and the anti-resonances.

We also note that at the edges of quantized conductance plateau
(edges of sub-bands also), such as at the points  D, E, and F in
Fig. \ref{PenEG}, the conductance has a kink-like or a dip-like
behavior. This enhanced scattering was also observed in disordered
graphene with non-topological impurities \cite{Zhang07}, and can be
attributed to extreme level-broadening (or velocity renormalization)
due to van Hove singularity at the sub-band edges \cite{QTD}. As can
be seen from Fig. \ref{PenEG}, this enhanced scattering is also
accompanied by a (qusi-) loop current on the defect, albeit with
different behavior from the purely topological localization
previously discussed: the directions of $\langle
i_{L}\rangle_{\text{pentagon}}$ and/or $\langle
m\rangle_{\text{pentagon}}$ are singly-polarized near the
conductance dip or peak, while the magnitude has a sharp peak.

\begin{figure} [t]
\begin{center}
\includegraphics[width=0.5\textwidth]{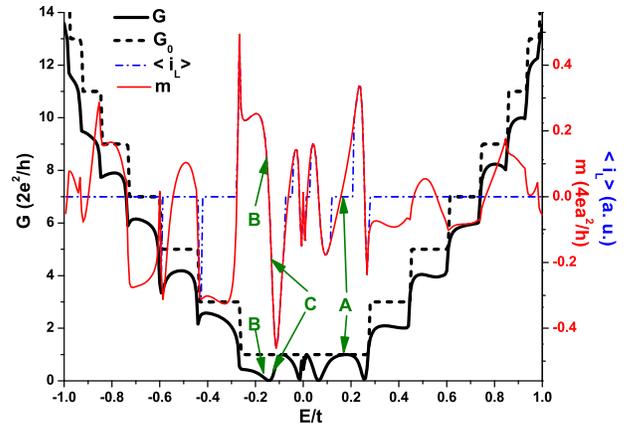}
\end{center}
\caption{(color online) Conductance of the topologically disordered
sample with a heptagon $G$ (black solid), averaged loop current
$\langle
 i_{L}\rangle$ (blue dashed dot) and induced magnetic moment $m$ (red solid) of the
pentagon as functions of Fermi energy $E$. The conductance for a
perfect graphene sample $G_0$ (black dashed) with the same
dimensions is also plotted for reference.} \label{HepEG}
\end{figure}
\begin{figure} [t]
\begin{center}
\fbox{\includegraphics[scale=0.3]{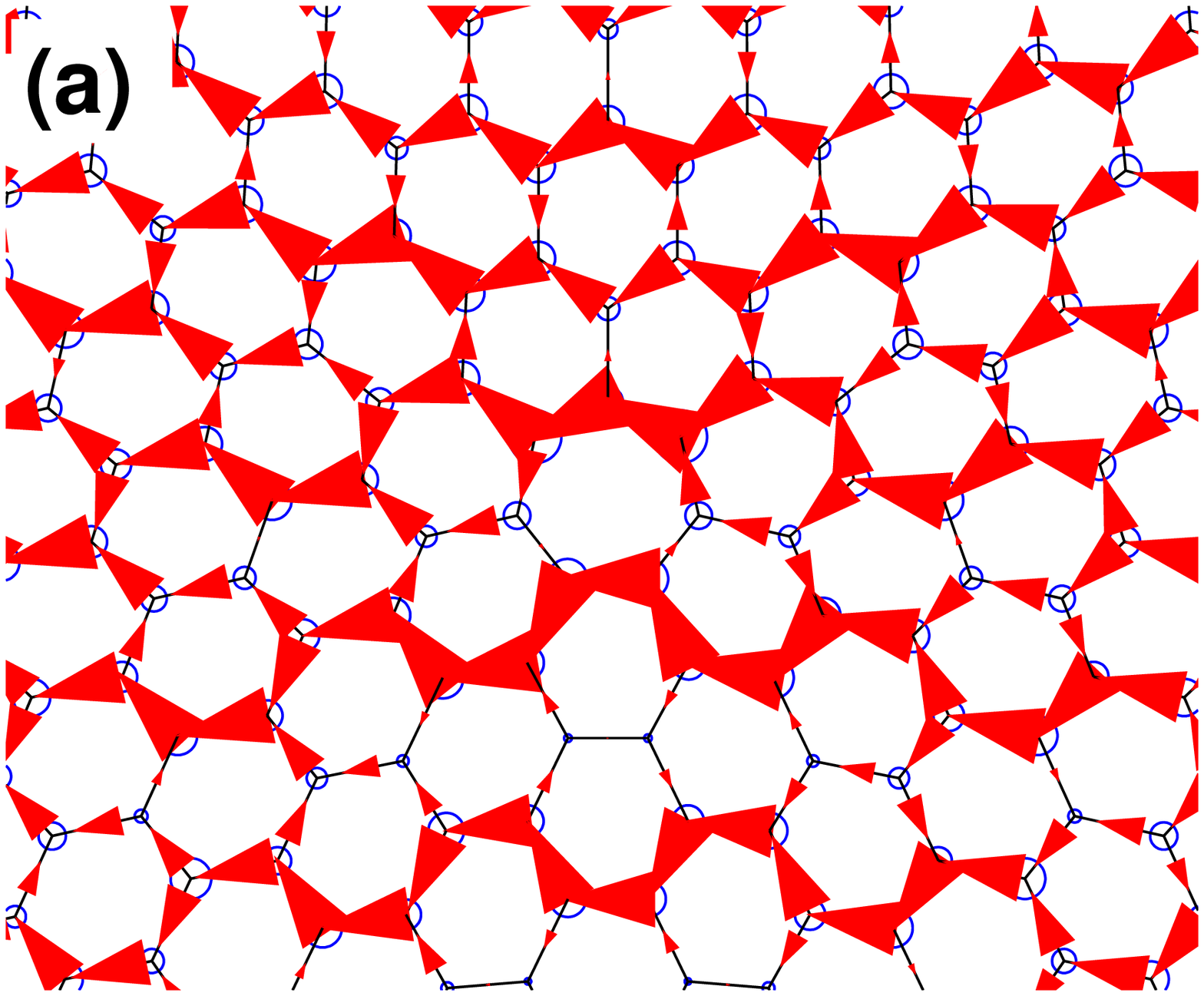}}
\fbox{\includegraphics[scale=0.3]{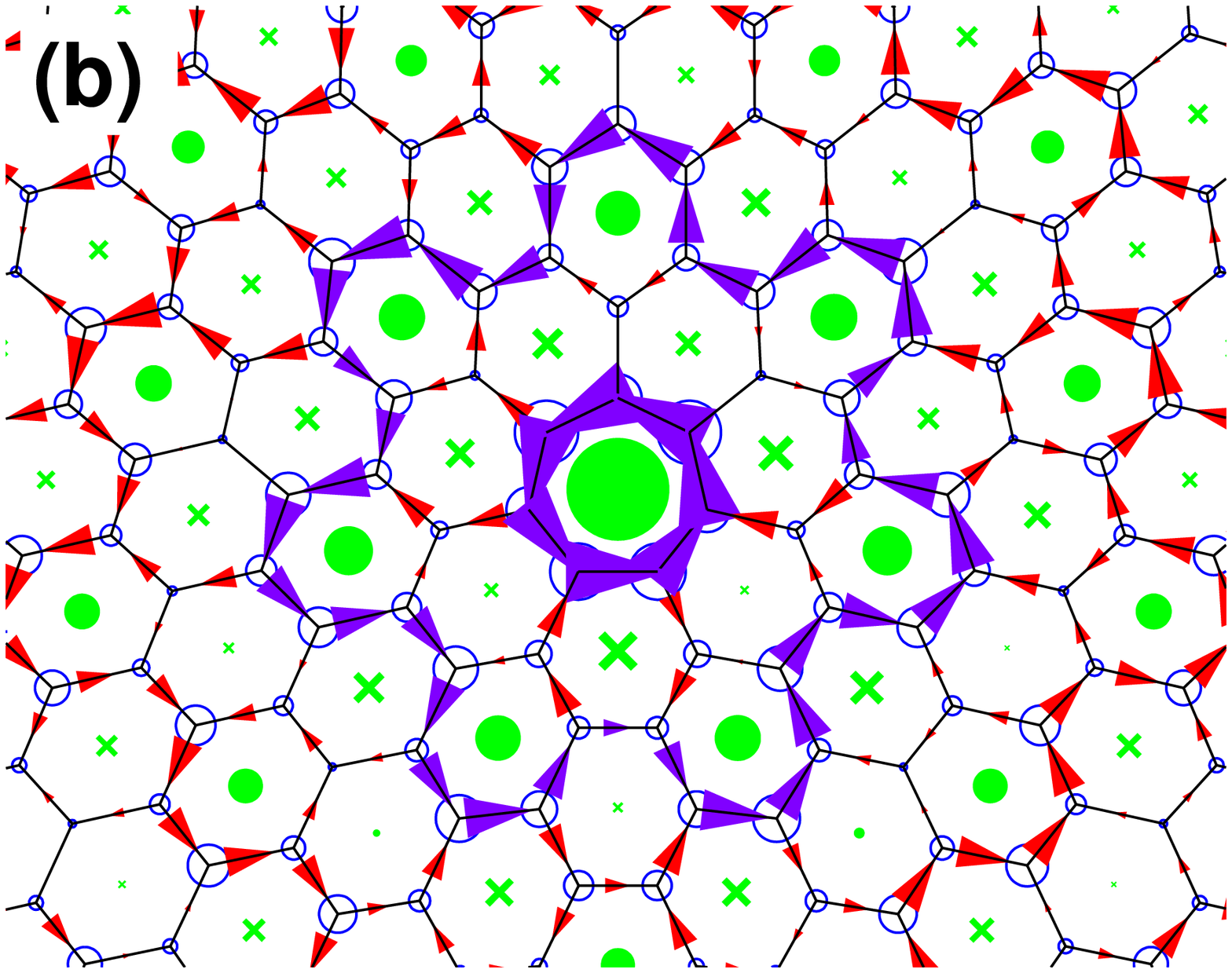}}
\fbox{\includegraphics[scale=0.3]{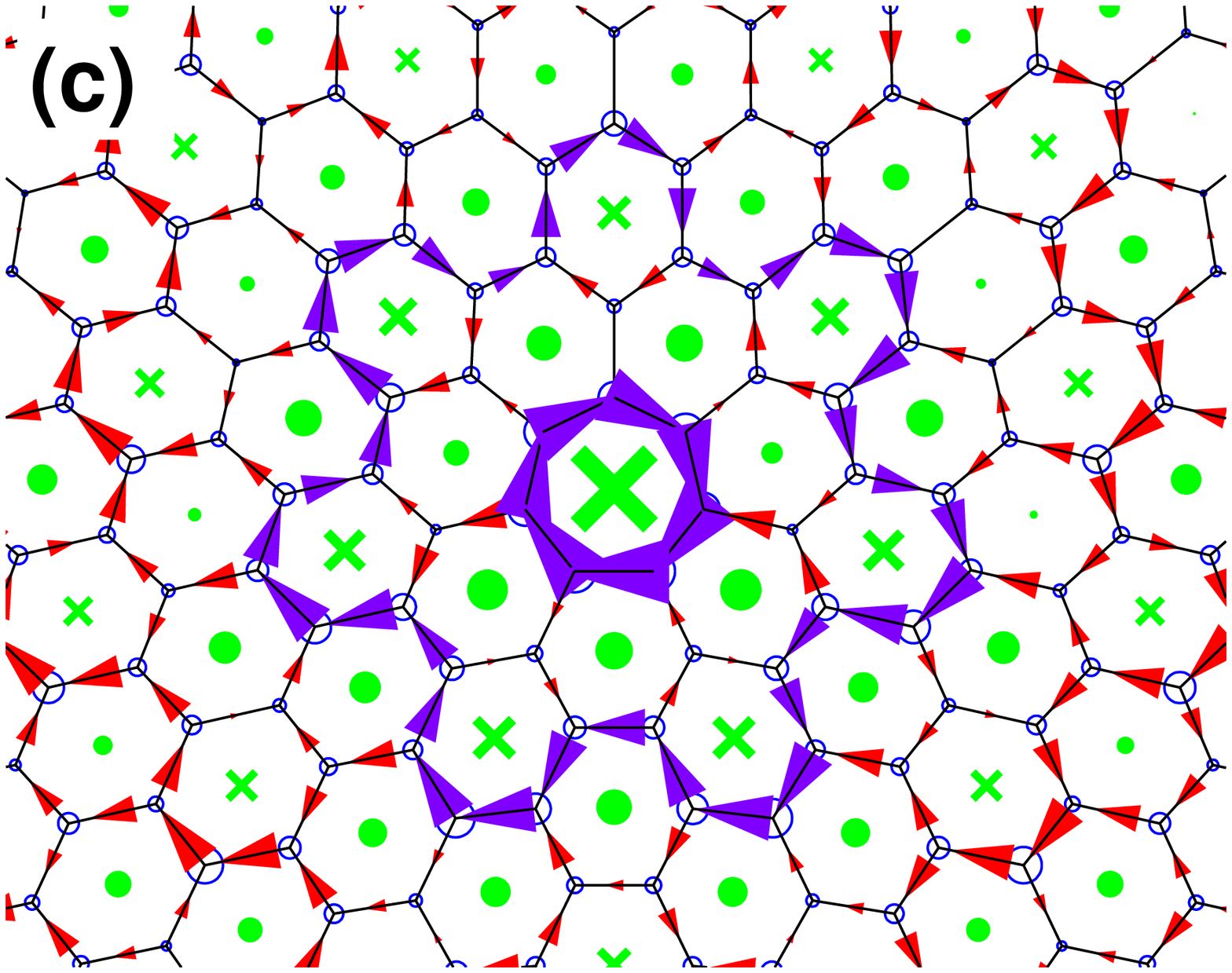}}
\end{center}
\caption{(color online) The spatial distribution of LDOS of the
sites (blue circles), local current density of the bonds (arrows)
and induced magnetic moment of carbon rings (green disks and
crosses) near the heptagon. The size of each symbol is proportional
to the local value of the corresponding quantity. (a) $E=0.17t$,
corresponding to point A in Fig. \ref{HepEG}, far away from the
anti-resonance. (b) $E=-0.15t$, corresponding to point B in Fig.
\ref{HepEG}, near the anti-resonance. (c) $E=-0.12t$, corresponding
to point C in Fig. \ref{HepEG}, on the other side of the
anti-resonance. Some loop currents around the pentagon are stressed
in purple arrows. Note that circular loop currents can also be
observed on some hexagons near the defect.} \label{Heptagon}
\end{figure}

The transport properties of graphene with a single heptagon are also
analyzed. Fig.\ref{HepEG} and Fig.\ref{Heptagon} are the results for
constructing a heptagonal defect in the center of a zigzag edge
graphene sheet with $30\times16$ sites. The quantum localizations,
vortex-like loop currents around the defect and reversal of the
chirality across the anti-resonance can also be observed, except the
pattern of loop currents now has seven-fold rotationally symmetry.

\begin{figure} [t]
\begin{center}
\includegraphics[width=0.5\textwidth]{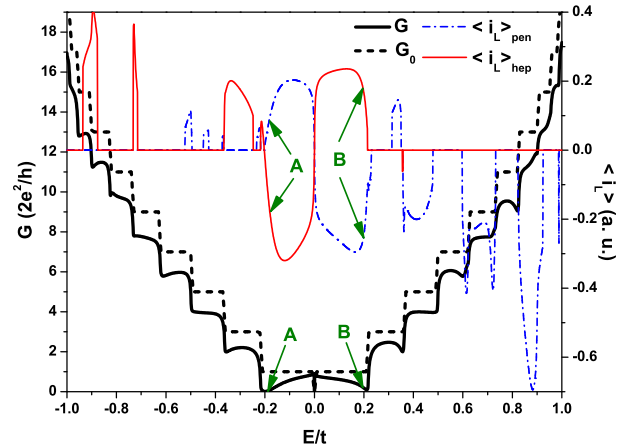}
\end{center}
\caption{(color online) Conductance $G$ (black solid) of the
topologically disordered sample with a pentagon-heptagon pair,
averaged loop current of the pentagon $\langle i_{L}\rangle_{pen}$
(blue dashed dot) and the heptagon $\langle i_{L}\rangle_{hep}$ (red
solid) as functions of Fermi energy $E$.} \label{PairEG}
\end{figure}

\begin{figure} [t]
\begin{center}
\fbox{\includegraphics[scale=0.3]{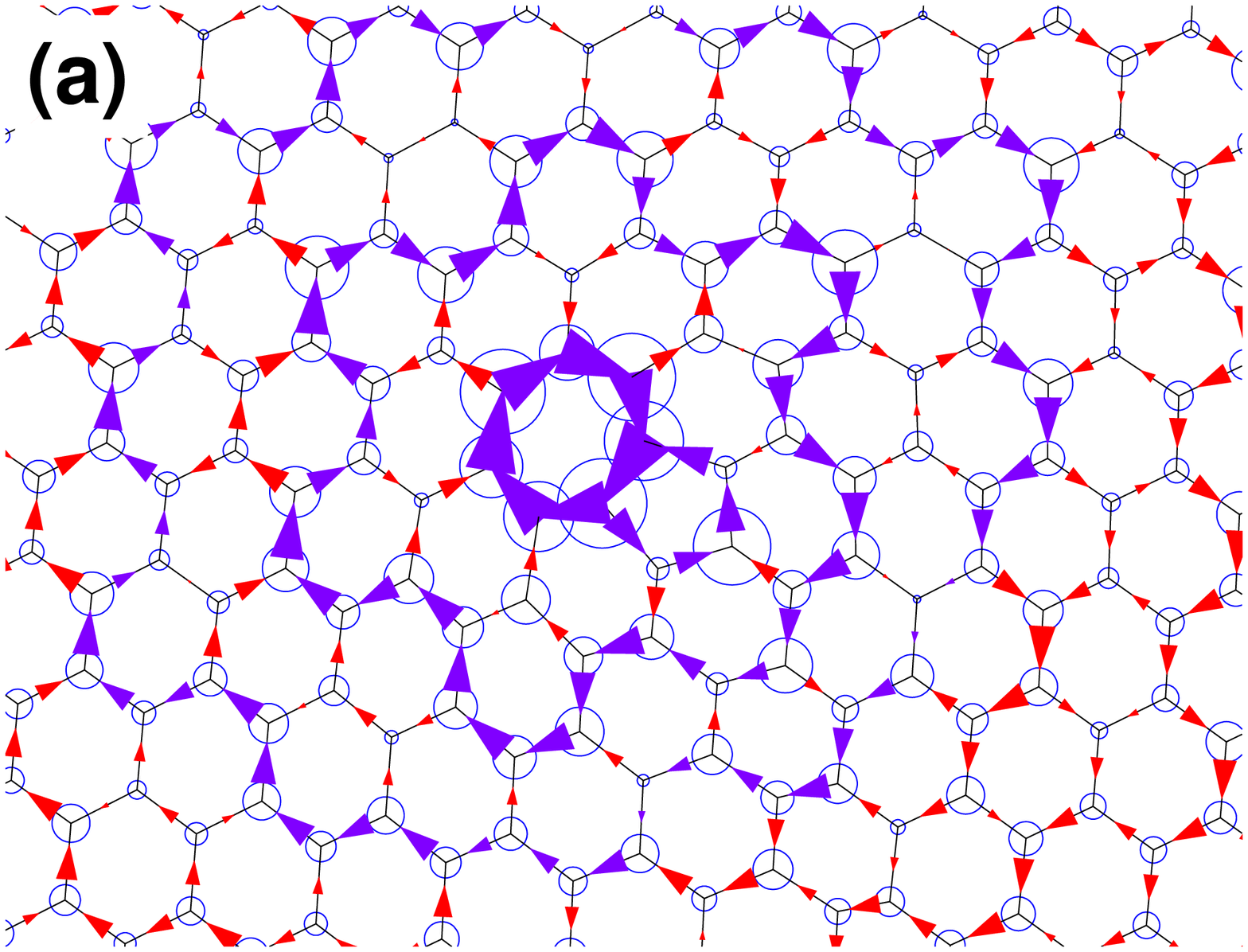}}
\fbox{\includegraphics[scale=0.3]{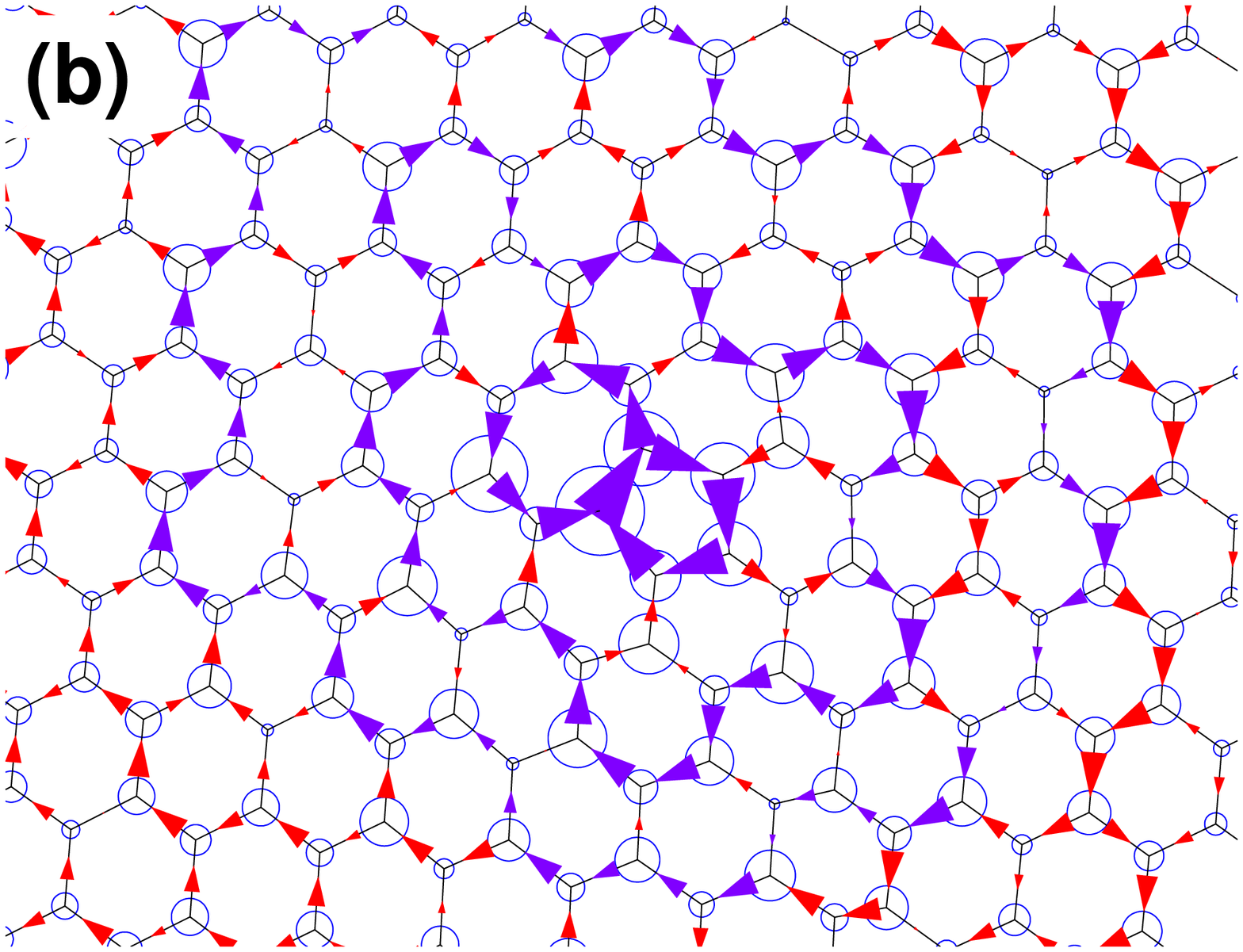}}
\end{center}
\caption{(color online) The spatial distribution of LDOS of the
sites (blue circles), local current density of the bonds (arrows)
near the pentagon-heptagon pair. (a) $E=-0.18t$, corresponding to
point A in Fig. \ref{PairEG}. (b) $E=0.2t$, corresponding to point B
in Fig. \ref{PairEG}. Some loop currents near the defect are plotted
in purple arrows.} \label{Pair}
\end{figure}

We now consider the pentagon-heptagon at the center of graphene.
With almost zero curvature, this is believed to be a stable
configuration in carbon-related materials. In Fig. \ref{PairEG}, the
conductance $G$, the ALCs $\langle i_{L}\rangle_{\text{pentagon}}$
and $\langle i_{L}\rangle_{\text{heptagon}}$ are plotted. Within the
first conductance plateau, the anti-resonance only happens at the
center and the edges of the plateau, and $\langle
i_{L}\rangle_{\text{pentagon}}$ and $\langle
i_{L}\rangle_{\text{heptagon}}$ always point in opposite directions.
On the higher conductance plateaus, we observe different energy
dependence of the two loop currents: non-zero loop current on the
pentagon tends to appear for $E>0$ while it tends to appear for
$E<0$ on the heptagon. This can be seen more clearly when we compare
the local currents near the anti-resonance at negative (Fig.
\ref{Pair} (a)) and positive energy (Fig. \ref{Pair} (b)),
respectively. There are also vortex-like loop currents encircling
the defect. When $E<0$, the magnitude of loop current on the
heptagon is larger than that on the pentagon, and the loop currents
encircling the whole defect have the same direction with the
heptagon loop current(Fig. \ref{Pair} (a)). Viceversa, when $E>0$,
the magnitude of loop current on the pentagon is larger, and the
vortex-like loop currents have the same direction with the pentagon
loop current (Fig. \ref{Pair} (b)). All these observations lead to a
nontrivial conclusion that, when the pentagon-heptagon pair is
present, the pentagon tends to trap the electronic motion in the
positive energy region, while the heptagon is dominant in the
negative energy region. The LDOS has a large magnitude on the
defect, similar to the case of disordered carbon nanotubes
\cite{Louie00}.

The quantum localization is realized only in the first quantized
channel as the conductance never vanishes completely in higher
conductance plateau (C and P in Fig. \ref{PenEG}). In a graphene
narrow ribbon with zigzag edges, the first quantized level is
different from higher levels due to the presence of edge
states\cite{Akh07}. In the first quantized level, for a given Fermi
level $E_F$, the currents carried in the states of one valley are
rectified currents and the directions are opposite in different
valleys. However, in the higher quantized levels, the states in each
valley carry current in both directions. In the presence of the
topological defects, the non-Abelian gauge potential scatters the
states in one valley to states in the other valley. In the case of
the first quantized level, the scattering can block the transport
current and create the loop current, since the process reverse the
current completely, while this is not the case in the high quantized
levels.

It is interesting to compare our results with those of other carbon
related structures. In Ref. \cite{Naka01}, a single C$_{60}$
molecule with discrete energy levels is connected to 1D leads, where
pentagonal loop currents were discovered. While, in our work, the
sample along with the leads is two-dimensional (2D) with continuous
energy bands. Therefore, the role of the degenerate \emph{resonance}
level in Ref. \cite{Naka01} is quite analogous to that of the
\emph{anti-resonance} level here, where the loop currents reverse
their directions. The perfect symmetry may be responsible for a much
larger loop current in C$_{60}$ (several tens of $I_{SD}$) than in
disordered graphene (few times of $I_{SD}$). Electronic transport
through topologically disordered carbon nanotubes (CNTs) has been
investigated \cite{Louie00}. But the topological defect therein
should possess certain configuration to satisfy the periodic
boundary condition in the transverse direction. Moreover, the
circular motions of the electron was verified to play important
roles in the transport, which is lacked in finite graphene ribbon
with fixed boundary condition.

Before arriving at final conclusions, it is important to remark
that, the understanding of bounding states associated with a single
topological defect discussed here can be a starting point for
investigating the effect of multiple topological defects. The
interferences between the bounding states of the defects may give
rise to some rich phenomena.

In conclusion, we have performed a numerical calculations of
conductance and local currents for topologically disordered graphene
with one pentagon, heptagon or pentagon-heptagon pair. A microscopic
understanding of the conductance reduction is obtained. The strong
scattering in these systems is always accompanied by (quasi-) loop
currents around the defects. The chirality of the loop currents can
be controlled by the bias voltage as well as the gate voltage near
the discrete Fermi energies where quantum localization takes place.
The magnetic moments generated by the loop currents can be measured
in experiments, such as scanning tunneling microscope(STM) and Kerr
effects. In the presence of the pentagon-heptagon pair defect, the
pentagon (heptagon) is more likely to trap electrons with positive
(negative) energy.

We thank Prof. F. Liu for discussions. This work is supported by RGC
CERG603904, NSF of China under grant 90406017, 60525417,
10740420252, the NKBRSF of China under Grant 2005CB724508 and
2006CB921400. JPH is supported by the US-NSF (Grant No.
PHY-0603759). XCX is supported by US-DOE and NSF.

\end{document}